\pdfoutput=1
\documentclass[prd,twocolumn,preprintnumbers,nofootinbib]{revtex4}
\usepackage{graphicx}
\usepackage{epsfig}
\usepackage{bm}
\usepackage{latexsym,amssymb,amsmath,amsfonts,amssymb,txfonts,wasysym,float}
\usepackage{scrextend}
\usepackage{mathrsfs}
\usepackage{color}

\usepackage{enumitem}

\newcommand{\postscript}[2]{\setlength{\epsfxsize}{#2\hsize}
   \centerline{\epsfbox{#1}}}

\usepackage[usenames,dvipsnames]{xcolor}
\definecolor{orange}{cmyk}{0,0.5,1,0}
\definecolor{rossoCP3}{cmyk}{0,.88,.77,.40}
\definecolor{graa}{rgb}{0.8,0.8,0.8}
\definecolor{blaa}{rgb}{0.2,0.2,0.6}

\begin{document}

\title{\color{rossoCP3} Decaying dark matter, the $\bm{H_0}$ tension,
  and the lithium problem}

\author{Luis A. Anchordoqui}
\affiliation{Physics Department, Herbert H. Lehman College and Graduate School, The City University of New York\\ 250 Bedford Park Boulevard West, Bronx, New York 10468-1589, USA}

\date{October 2020} 

\begin{abstract}
  \noindent It has long been known that the sharpened tension between
  the observed and inferred values of the Hubble constant $H_0$ can be
  alleviated if a fraction of dark matter particles of type $\chi$
  were produced non-thermally in association with photons $\gamma$
  through the decays of a heavy and relatively long-lived state, {\it
    viz.}  $X \to \chi \gamma$. It was recently proposed that this
  model can also resolve the longstanding lithium (a.k.a. $^7$Li) problem if
  \mbox{$M= 4~{\rm MeV}$} and \mbox{$m = 0.04~{\rm keV}$,} where $M$ and $m$ are
  respectively the masses of $X$ and $\chi$. We confront this proposal
  with experiment and demonstrate that cold dark matter decaying
  before recombination cannot resolve the $H_0$ problem. Moreover, we show that the best case scenario for alleviating the $H_0$ tension within the context
  of  cold dark matter decaying before recombination arises when the
  particles decay exclusively into dark radiation, while leaving
  completely unmodified the production of light elements.  To this end we calculate the general functional form
  describing 
  the number of equivalent light neutrino species $\Delta N_{\rm eff}$
  carried by $\chi$. We show that to resolve the $H_0$ tension at the
  $1\sigma$ level a 55\% correction in $m$ is needed and
  that the required $\Delta N_{\rm eff}$ is excluded at 95\% CL
  by Planck data. We argue in favor of a more complex model of
  dynamical dark matter to relax the $H_0$ tension.
\end{abstract}
\maketitle

\section{Introduction}

Over the past decade cosmological parameters have been measured to
unprecedented precision. The most reliable measurement of the Hubble
constant $H_0 = 74.03 \pm 1.42~{\rm km/s/Mpc}$ comes from HST observations  of
Cepheid variables in the host of recent, nearby type Ia supernova to
build a 3-rung distance ladder~\cite{Riess:2019cxk}. Besides, a prediction of $H_0$ can be obtained from the
sound horizon observed from the cosmic microwave background (CMB). A fit to
data from the Planck mission, under the assumption of a flat $\Lambda$
cold dark matter ($\Lambda$CDM) cosmological model leads to $H_0 = 67.27 \pm 0.60~{\rm km/s/Mpc}$~\cite{Aghanim:2018eyx}. These
two $H_0$ values are discrepant by about $4.4\sigma$, which gives rise
to the so-called $H_0$ tension~\cite{Verde:2019ivm,DiValentino:2020zio}.

Although big bang nucleosynthesis (BBN) has played a central role in the
development of precision cosmology, it appears the theory has hitherto failed to explain the amount of cosmic
lithium~\cite{Fields:2011zzb}. Indeed, while BBN predictions match the
observed primordial deuterium (D)
and helium ($^3$He, $^4$He) abundances, the theory seems to over-predict the abundance of
primordial lithium ($^7$Li) by about a factor of three. This $4 - 5\sigma$
mismatch constitutes the so-called ``cosmic $^7$Li problem.''

Both short-lived ($\tau \ll t_{\rm LS}$)  and long-lived ($\tau
\gg t_{\rm LS}$) dark matter
particles decaying into dark radiation provide  promising scenarii to
tackle the tension on the expansion rate (where $\tau$ is the particle's lifetime and $t_{\rm LS}$ denotes the time of last scattering)~\cite{Hooper:2011aj,GonzalezGarcia:2012yq,Berezhiani:2015yta,Vattis:2019efj,Blinov:2020uvz}. To understand why this is so,
we begin by noting that the CMB anisotropy power spectrum tightly constrains the angular size
of the sound horizon at recombination $\theta_*$, which in a flat universe is given by the
ratio of the comoving sound horizon to the
comoving angular diameter distance to last-scattering surface: $\theta_* = r_s(z_{\rm LS})/D_M(z_{\rm LS})$.
The comoving linear size of the of the sound horizon and the comoving angular
diameter distance are linked to the expansion history of the universe
via $r_s(z) = \int_z^\infty c_s(z') \, dz'/H(z')$  and $D_M(z) =  \int_0^z  dz'/
H(z')$, respectively, with $c_s$ the speed of sound and $H(z)$ the Hubble
parameter at redshift $z$~\cite{Aghanim:2018eyx}.

For $\tau \ll t_{\rm LS}$, matter is depleted into radiation at
redshifts $z > z_{\rm LS}$. The main effect of adding radiation
density to the early universe is to increase the expansion rate
$H(z)$, which in turn reduces the sound horizon during the era leading
up to recombination. Since the location of the acoustic peak is
accurately measured, $\theta_*$ must be kept fixed. This can be
accomplished, e.g., by simultaneously increasing $H_0$. However, adding
radiation density to the early universe also alters the damping scale
$\theta_D$ of the CMB power spectrum, with $\theta_D/\theta_* \propto
\sqrt{H(z_{\rm LS})}$~\cite{Hou:2011ec}. Increasing $H(z_{\rm LS})$ at fixed $\theta_*$
leads to an increase of $\theta_D$, and so  the damping kicks in at
larger scales reducing the power in the damping tail. An accurate
measurement of the small-scale CMB anisotropies thus 
constrains the fraction of short-lived CDM decaying into radiation.

For  $\tau
\gg t_{\rm LS}$, matter is depleted into radiation at redshifts $z <
z_{\rm LS}$. Therefore, the sound horizon is regulated by the state of the
universe prior to last scattering, and so the value of $r_s (z_{\rm LS})$ does not differ appreciably from that obtained assuming
$\Lambda$CDM for the same choice of cosmological parameters. However,
as a result of the cosmic expansion the radiation density 
decreases more rapidly than the matter density, and therefore the
local expansion rate of the universe is lower in the late-time decaying dark matter scenario than it would have been in 
$\Lambda$CDM with the same value of $H(z_{\rm LS})$. Now, a
consistently lower value of $H(z)$ at low redshifts leads to a larger
value of $D_M(z_{\rm LS})$, which in turn would result in a smaller
value of $\theta_*$. Again, the location of the acoustic peak is
accurately measured, and so $\theta_*$ must be kept fixed. This can be
accomplished, e.g., by increasing the dark energy density $\Omega_\Lambda$. As a consequence of the larger $\Omega_\Lambda$ the matter-dark energy equality is
shifted to earlier times than it otherwise would in $\Lambda$CDM, yielding a
larger value of $H_0$. Remarkably, such late-time decays can also resolve
the growing tension between the cosmological and local determination
of $S_8 \equiv \sigma_8 \sqrt{\Omega_m({\rm today})/0.3}$, which quantifies the
rms density fluctuations when smoothed with a top-hat filter of radius $8
h^{-1}/{\rm Mpc}$~($\equiv \sigma_8$) as a function of  the present
day value of the nonrelativistic matter density parameter $\Omega_m({\rm today})$, where $h$ is the dimensionless Hubble constant~\cite{Enqvist:2015ara,Abellan:2020pmw,DiValentino:2020vvd}.

It was recently proposed that if a fraction of dark matter particles
$\chi$ were produced non-thermally in association with photons through
the decays of a heavy and relatively long-lived  state, both the $H_0$
and $^7$Li  problems can be simultaneously
resolved~\cite{Alcaniz:2019kah} (see~\cite{DiBari:2013dna}, for a
similar proposal). In this paper we confront this proposal with
experiment and demonstrate that short-lived CDM cannot
resolve the $H_0$ problem. Moreover, we show that the best case scenario for
alleviating the $H_0$ tension within the context of short-lived CDM
arises when the particles decay exclusively into dark radiation, while
leaving completely unmodified the production of light elements.  The
paper is structured as follows. In Sec.~\ref{sec:2} we calculate the
general functional form describing the number of equivalent light
neutrino species $\Delta N_{\rm eff}$ carried by $\chi$. We show that
to resolve the $H_0$ tension at the $1\sigma$ level a 55\% correction
in the mass of $\chi$ is needed and that the required
$\Delta N_{\rm eff}$ is excluded at 95\% CL by Planck data. In
Sec.~\ref{sec:3}  we argue in favor of a more complex hidden sector that could
combine ``early-time'' and ``late-time'' decaying dark matter solutions of the $H_0$
tension. The paper wraps up with some conclusions in Sec.~\ref{sec:4}.

\section{Constraints on short-lived CDM}

\label{sec:2}

\begin{figure*}[tb] 
    \postscript{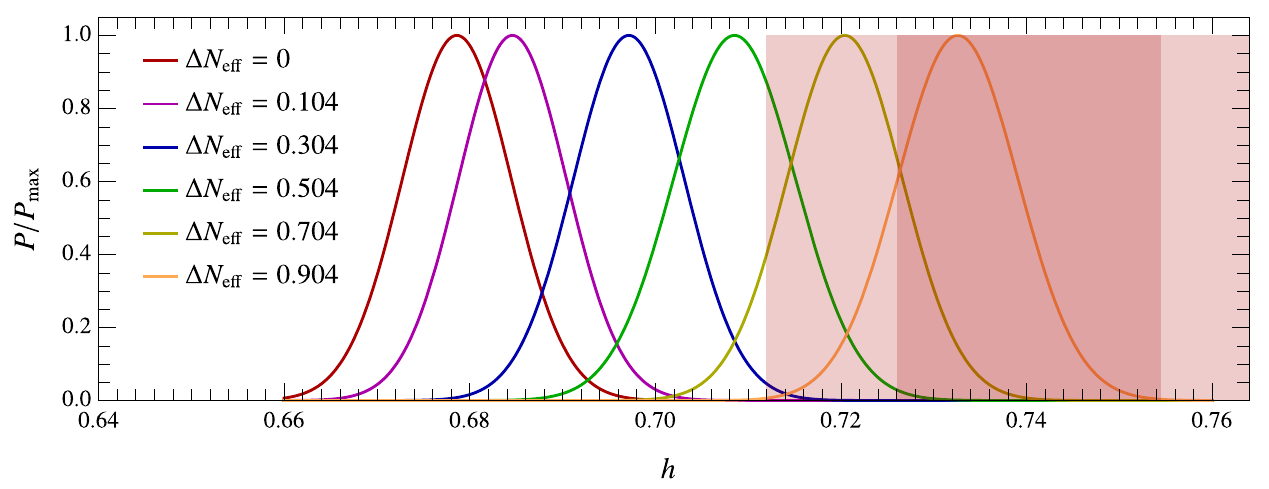}{0.99} 
    \caption{Rescaled posterior distributions of $H_0 = 100~h~{\rm km/s/Mpc}$ (due to marginalization over additional free parameters) with different
      choices of $\Delta N_{\rm eff}$ from the 7 parameter fit 
      of~\cite{Vagnozzi:2019ezj}. The
      shaded areas indicate the $1\sigma$ and $2\sigma$ regions as
      determined with HST observations~\cite{Riess:2019cxk}.
\label{fig:uno}}
\end{figure*}

Following~\cite{Alcaniz:2019kah} we assume that of the total dark
matter (DM) density around today, $\Omega_{\rm DM}({\rm today})$, a
small fraction, $f = \Omega_\chi ({\rm today})/\Omega_{\rm DM}({\rm today})$, is of particles of
type $\chi$, produced via decay of a heavy relic $X$ with mass $M$ and
lifetime $\tau$: $X \rightarrow \chi \gamma$.  At any time after the decay of $X$
the total DM energy density is found to be
\begin{equation}
\rho_{\rm DM} (t) = \frac{m \ n_\chi ({\rm today})}{a^3(t)} \ \gamma (t) +
(1-f) \ \rho_c \ \frac{\Omega_{\rm DM} ({\rm today})}{a^3(t)} \,,
\end{equation}
where $m$ is the mass of the $\chi$, $\gamma (t)$ its Lorentz
boost, $n_\chi(t)$ its number density, $\rho_c$ the critical density, and $a(t)$ is the expansion scale factor normalized by $a({\rm today})
= 1$. In the center-of-mass frame of $X$ (this should also be a good approximation of any frame as we assume that $X$ is nonrelativistic so its mass energy dominates) we have
\begin{equation}
M = E_0 + p_0 = \sqrt{p_0^2 + m^2} + p_0 \Rightarrow p_0 = \frac{M^2 -
  m^2}{2 M} \,,
\end{equation}
where $E_0 = \gamma (\tau) m$  is the initial energy of the particle
$\chi$ at $z_{\rm decay}$ and $p_0$ its momentum, with
\begin{equation}
  \gamma (\tau) = \frac{E_0}{m} = \frac{M}{2 m} + \frac{m}{2 M} \, .
  \end{equation}
Because of the cosmic expansion any particle with momentum $p$ becomes
redshifted at a rate $p(t) = p_0 a(\tau)/a(t)$. Using the relation
$E^2 - p^2 = m^2$, which holds in the Robertson-Walker
metric~\cite{Kolb:1988aj}, we find that the particle's energy gets redshifted according to
\begin{equation}
  E^2 (t) = p_0^2 \ \frac{a(\tau)^2}{a (t)^2} + m^2 \, .
\end{equation}
The scale factor dependence on the Lorentz boost is found to be
\begin{eqnarray}
  \gamma(t) & = & \sqrt{1 + \left[\frac{a(\tau)}{a(t)}\right]^2
                       \left(\frac{p_0}{m} \right)^2} 
  =  \sqrt{ 1 + \left[\frac{a(\tau)}{a(t)}\right]^2 \left(\frac{E_0^2 -
  m^2}{m^2}\right)} \nonumber \\
               & = & \sqrt{1 + \left[\frac{a(\tau)}{a(t)} \right]^2
                     \left[\gamma^2(\tau) - 1 \right]} \, .
\label{gammacuadrado}                     
\end{eqnarray}
Expansion of the square root in
(\ref{gammacuadrado}) leads to
\begin{equation}
\gamma(t)   \approx  1 + \frac{1}{2} \left[\frac{a
  (\tau)}{a(t)}\right]^2 \left[\gamma^2 (\tau) -1 \right] - \frac{1}{8} \left[
\frac{a (\tau)}{a(t)} \right]^4  \left[ \gamma^2 (\tau) -1 \right]^2 
+ \cdots .
\label{expansion}
\end{equation}
Now,  $\Omega_\chi ({\rm today}) =  m \ n_\chi ({\rm today})/\rho_c$, because the $\chi$ is non-relativistic.  To obtain
such a non-relativistic limit we demand the magnitude of the second
term in the expansion of (\ref{expansion}) to be greater than the
third term, which results in $[a (\tau)/a(t)]^2 [\gamma^2 (\tau) -1] <
4$. Contrariwise, by this criteria the particle $X$ is relativistic if
$\gamma (t) > \sqrt{5}$. A point worth noting at
this juncture is that the general functional form of the Lorentz boost given
in (\ref{gammacuadrado}) and its expansion given in (\ref{expansion})
are substantially different from the approximate expression of $\gamma (t)$ given in Eq.~(3)
of~\cite{Hooper:2011aj}. However, for $\gamma^2(\tau) \gg 1$ and $\gamma^2(\tau) [a
(\tau)/a(t)]^2 \gg 1$, both expressions give similar Lorentz factors.

The total ``dark'' relativistic
energy density (including the three left-handed neutrinos of the
Standard Model) is
usually characterized by the number of ``equivalent'' light
neutrino species, $N_{\rm eff} \equiv (\rho_{\rm R} -
\rho_\gamma)/\rho_{\nu_L}$,  in units
of the density of a single Weyl neutrino $\rho_{\nu_L}$, where $\rho_\gamma$ is the energy density of
photons, and $\rho_{\rm R}$ is the total
energy density in relativistic
particles~\cite{Steigman:1977kc}. Following~\cite{Alcaniz:2019kah}, 
we obtain the $\chi$ contribution to $N_{\rm eff}$ at the time of
matter-radiation equality  assuming that the $\chi$ decouples from the plasma prior to $\nu_L$ decoupling,
conserving the temperature ratio $T_\gamma/T_{\nu_L} = 11/4$ from $\Lambda$CDM cosmology,
\begin{eqnarray}
 \label{deltanX1}
  \Delta N_{{\rm eff}} & = & \frac{8}{7} \left(\frac{11}{4}
                             \right)^{4/3} \frac{\rho_\chi (t_{\rm
                             EQ})}{\rho_\gamma (t_{\rm EQ})} \nonumber \\
& = & 
                             \frac{8}{7} \left(\frac{11}{4}
                             \right)^{4/3} \frac{\Omega_{\rm
                           DM}({\rm today})}{\Omega_\gamma({\rm today})} \ a(t_{\rm EQ}) \ f \
      \gamma(t_{\rm EQ} ) \nonumber \\
& = & 
                            \frac{8}{7} \left(\frac{11}{4}
                             \right)^{4/3} \frac{\Omega_{\rm
      DM} ({\rm today})}{\Omega_\gamma({\rm today})} \ a(t_{\rm EQ})
      f \nonumber \\ 
 & \times &      \sqrt{1 + \left[\frac{a(\tau)}{a(t_{\rm EQ})} \right]^2
                     \left[\gamma^2(\tau) - 1 \right]}, 
\end{eqnarray}
where $\rho_\gamma (t_{\rm EQ}) = \rho_c \Omega_\gamma ({\rm today})/a^4(t_{\rm
  EQ})$ and the factor of $8/7$ is due to the difference between the Fermi and
Bose integrals~\cite{Anchordoqui:2013wwa}. Note that the functional
form of (\ref{deltanX1}) is different from that in Eq.~(5)
of~\cite{Hooper:2011aj}; the latter was adopted
in the recent study of~\cite{Alcaniz:2019kah}.  For $\gamma^2(\tau) \gg 1$ and $\gamma^2(\tau) [a
(\tau)/a(t)]^2 \gg 1$, both (\ref{deltanX1}) and Eq.~(5) of~\cite{Hooper:2011aj} give similar contributions to
$\Delta N_{\rm eff}$.

Model considerations set bounds on free parameters. On the one hand,
consistency with large-scale structure observations implies $f \alt
0.01$~\cite{Hooper:2011aj}. On the other hand, the decay of the $X$'s could significantly 
alter the light element abundances synthesized during BBN. Of
particular interest here, the threshold energy of the photon for the
process $^7$Be($\gamma$,$^3$He)$^4$He is $E^{\rm th}_{^7{\rm Be}}
\simeq 1.59~{\rm MeV}$, which is lower than that of the
photodissociation of D ($E^{\rm th}_{\rm D} \simeq 2.22~{\rm MeV}$)
and $^4{\rm He}$ ($E^{\rm th}_{^4{\rm He}} \sim 20~{\rm MeV}$). Hence, if the energy of the injected photons is in the range 
$E^{\rm th}_{^7{\rm Be}} < E_{\gamma} < E^{\rm th}_{\rm D}$, the photodissociation of $^7$Be could take
place to solve the $^7$Li problem without significantly affecting the
abundances of other light elements~\cite{Poulin:2015woa}. With this in mind, to destroy enough $^7$Li without affecting the
abundance of other elements, we set $f \simeq 0.01$ and must  fine-tune simultaneously
the $X$-lifetime $\tau \simeq 2 \times 10^{-4}~{\rm s}$~\cite{Alcaniz:2019kah} and 
the initial electromagnetic energy release in each $X$ decay
$E^{\rm th}_{^7{\rm Be}} < E_{\gamma} = ({M}^2 - m^2)/(2 M) < E^{\rm th}_{\rm D}$. For $M \gg m$, the latter leads to $M \simeq 4~{\rm MeV}$.

The correlation between $H_0$ and $N_{\rm eff}$ has been estimated numerically
\begin{equation}
  \Delta H_0 = H_0 - \left. H_0 \right|_{\Lambda {\rm CDM}} \approx 6.2 \
  \Delta N_{\rm eff} \,,
  \label{H0}
\end{equation}
where  $H_0|_{\Lambda {\rm CDM}}$ is the value of $H_0$ inferred
within $\Lambda$CDM~\cite{Vagnozzi:2019ezj}.\footnote{We note in
  passing that the relation (\ref{H0}) has been derived on the basis
  of the Planck
  2015 TT + lowP + BAO + Pantheon dataset
  combination~\cite{Ade:2015xua,Scolnic:2017caz}. Should we instead
  have used Planck 2018 TTTEEE + lowE + BAO + Pantheon~\cite{Aghanim:2018eyx}, the proportionality
  constant would turn out to be 5.9 rather 6.2~\cite{Sunny}. This is because
  small-scale polarization data comes into play making it  even harder
  to accommodate high $H_0$ values with $\Delta N_{\rm eff}$. Herein we remain
  conservative in our calculations and adopt the value given in~\cite{Vagnozzi:2019ezj}.} The rescaled posterior
distributions of $H_0$ from the a parameter fit with different choices
of $\Delta N_{\rm eff}$ are displayed in
Fig.~\ref{fig:uno}. To accommodate the $H_0$
tension at  $1\sigma$ level, we require $\Delta H_0 \approx 4.7$, which implies
$\Delta N_{\rm eff} \approx 0.76$.  The 95\% CL bound
on the extra equivalent neutrino species derived from a combination of CMB, BAO, and BBN observations is
$\Delta N_{\rm eff} < 0.214$~\cite{Aghanim:2018eyx}. This
  limit combines the helium measurements of~\cite{Aver:2015iza,Peimbert:2016bdg} with the latest
  deuterium abundance measurements of~\cite{Cooke:2017cwo} using the
  the \texttt{PArthENoPE} code~\cite{Pisanti:2007hk} 
considering $D(p,\gamma)^3{\rm He}$ reaction rates
from~\cite{Marcucci:2015yla}. Note that even considering the most
optimistic helium abundance measurement of~\cite{Izotov:2014fga} in place
of~\cite{Aver:2015iza,Peimbert:2016bdg}, the 95\% CL bound $\Delta N_{\rm eff} < 0.544$~\cite{Aghanim:2018eyx} still precludes accommodating the HST observations within $1\sigma$.

Following~\cite{Hooper:2011aj}, we take a radiation like scale factor
evolution $a(t) \propto t^{1/2}$, $\Omega_ {\rm DM} ({\rm today}) \approx 0.227$,
$\Omega_{\gamma} ({\rm today}) \approx 0.0000484$, $a(t_{\rm EQ}) \approx 3 \times
10^{-4}$, and $a(\tau)/a(t_{\rm EQ}) = 7.8 \times 10^{-4}
\sqrt{10^{-6} \, \tau/{\rm s}}$. Substituting these figures into (\ref{deltanX1})
while demanding the constraint $\Delta N_{\rm eff} \approx 0.76$  via (\ref{H0}) leads to  $m \simeq
0.018~{\rm keV}$.

If $m =0.07~{\rm keV}$ the contribution to
$\Delta N_{\rm eff} \simeq 0.2$ is consistent with the upper 
limit reported by the Planck Collaboration~\cite{Aghanim:2018eyx} and the $X \rightarrow \chi
+ \gamma$ decays still produce enough electromagnetic energy~\cite{Hooper:2011aj}
\begin{equation}
\varepsilon_\gamma \approx 1.5 \times 10^{-6} \ f \ \left(\frac{M}{m} -
  \frac{m}{M}\right)~{\rm MeV}    
\end{equation}
to dilute the $^{7}$Li~\cite{Poulin:2015woa}. However, a recent study of the
photodissociation of light elements in the early universe indicates
that the bound on the electromagnetic energy in 2~MeV photons that
can be injected during BBN is $\varepsilon_\gamma < 10^{-4}~{\rm
  MeV}$~\cite{Kawasaki:2020qxm}. This further constrains the
contribution to $\Delta N_{\rm eff}$ via $X \rightarrow \chi 
\gamma$. On the other hand, if the $X$ particle decays exclusively
into dark radiation the contribution of this process to $\Delta N_{\rm eff}$
is only limited by $\Delta N_{\rm eff} < 0.214$~\cite{Aghanim:2018eyx}, allowing for  $\Delta H_0 \sim 1.3$. 

\section{$\bm{H_0}$ tension and
   dynamical dark matter}

\label{sec:3} 

Dynamical dark
matter (DDM)~\cite{Dienes:2011ja} provides a self-sustaining framework to unify short-lived
and long-lived decaying dark matter models. In the DDM framework the
requirement of dark-matter stability is swapped by a balancing of
lifetimes against cosmological abundances across an ensemble of
individual dark-matter components with different masses, lifetimes,
and abundances. This DDM ensemble collectively
plays the role of the dark-matter ``candidate'' while collectively
describing the observed dark-matter abundance.

Following~\cite{Desai:2019pvs}, we consider an ensemble comprising a
large number $N$ of individual constituent particle species $X_n$, which decay via $X_n \to \psi \bar \psi$, where  $\psi$ is a massless dark-sector particle which
behaves as dark radiation and
the index $n = 1,2, \cdots, N$ labels the particles in order of
increasing mass $m_n$. The total decay widths $\Gamma_n$ scale across
the ensemble satisfying $\Gamma_1 < \Gamma_2 < \cdots < \Gamma_N$,
where $\Gamma_1$ is the decay width of the lightest particle in
the ensemble. The
initial abundances $\Omega_n (z_{\rm prod})$ are regulated by early universe
processes and fixed at $z_{\rm prod} \gg z_{\rm LS}$, with $t_{\rm prod} \ll
\tau_{N}$. 

 For simplicity, we first assume that all particles in
the ensemble are cold, in the sense that their equation-of-state
parameter may be taken to be $w_{n} \approx 0$ for all $t > t_{\rm
  prod}$. If we further assume that $X_{N-1}$ is almost stable ($\Gamma_{N-1} \ll H_0$), then the
DDM framework can accommodate short-lived or long-lived CDM models
decaying exclusively into dark radiation.

On the one hand, if 
$\Gamma_N \gtrsim 10^6~{\rm  Gyr}^{-1}$,  the presence of additional energy at around the time of matter-radiation
equality contributing to the value of $\Delta N_{\rm eff}$ constrains
the abundance of the of the dark radiation field $\Omega_\psi (z_{\rm
  LS}) \alt 0.1 \Omega_\gamma (z_{\rm LS})$,  
where we have used the 95\% CL upper limit $\Delta N_{\rm eff} <
0.214$~\cite{Aghanim:2018eyx}.

On the
other hand, if $\tau_N \agt t_{\rm LS}$, the decay of $X_N$  gives
late-time solutions to the $H_0$ problem. These long-live decaying CDM
models can be further sub-classified by means of the particle's decay
width:
\begin{itemize}[noitemsep,topsep=0pt]
\item $\Gamma_N \gtrsim H_0 \sim 0.7~{\rm Gyr}^{-1}$, in which most of the $X_N$-particles have disappeared by
  $z=3$;
  \item $\Gamma_N \lesssim H_0$, in which only a fraction of the $X_N$-particles
    had time to disappear.
\end{itemize}
The total initial CDM abundance of
the DDM ensemble $\sum_n \Omega_n (z_{\rm prod})$ is
essentially fixed by the requirement that at recombination  $\sum_n
\Omega_n(z_{\rm LS})$ 
accommodates the dark matter abundance $\Omega_{\rm DM} (z_{\rm LS})$
derived from Planck data. The initial fraction $F \equiv \Omega_N (z_{\rm prod
  })/\sum_n \Omega_n (z_{\rm prod})$ is
a free parameter of the ensemble.

A full resolution of the
Hubble tension within the
first sub-class requires $F \sim 0.1$~\cite{Berezhiani:2015yta}, with possible implications for
the flux of cosmic neutrinos detected by
IceCube~\cite{Anchordoqui:2015lqa}. However,  a large amount of $X_N$-decay significantly suppresses the
amount of DM at low redshifts and thus reduces the power of the CMB
lensing effect, which is at odds with Planck
observations~\cite{Chudaykin:2016yfk}. Indeed, CMB and BAO
data constrain the fraction of decaying DM,  $F < 0.01$~\cite{Poulin:2016nat,Chudaykin:2017ptd,Nygaard:2020sow}, and hence
disfavor a full resolution of the $H_0$ tension if $0.7 \alt  \Gamma_N/{\rm Gyr}^{-1}
\alt 10^6$.

An explanation of the Hubble tension within
the second sub-class allows for smaller values of $F$, but requires a
more complex structure of the ensemble with
two additional assumptions: {\it (1)}~ allow for intra-ensemble
decays, {\it viz.} decays of $X_N$ into final states
that include other, lighter $X_n$; {\it (2)} not all the $X_n$ in the
ensemble are cold particle species~\cite{Vattis:2019efj}. If this were the case then the massive daughter
particle could be born relativistic at $z_{\rm decay}$ when the expansion
rate is given by $H(z_{\rm decay})$, but behave like CDM as the
universe evolves, yielding a dynamic equation of state $w_{n}(z)$. Therefore, at any $z$ the collective behavior of
any $X_n$ which were born relativistic needs to be averaged over all particles that were born at higher
redshifts. This implies that the evolution of the energy density 
$\Omega_{X_n} (z)$ of these warm particles depends on the sum of all contributions of particles
born during the interval $0\leq z \leq z_{\rm decay}$, some of which were
born relativistic and redshifted away by $z = 0$ and some that are
born at late times but had no time to be redshifted. A related model
requires a DDM ensemble in which the decaying dark matter has an
appreciable free-streaming length, and therefore does not cluster on small scales
as CDM does, {\it viz.} $X_N$ is relativistic at
production~\cite{Blinov:2020uvz}. However, if we look underneath the seats, we find devils
carved: low multipoles amplitude of the CMB anisotropy power spectrum
severely constrains the feasibility of late-time decay models
(including those with intra-ensemble decays) to fully resolve to the $H_0$ tension~\cite{Haridasu:2020xaa,Clark:2020miy}.

The solution of this conundrum lies perhaps in a large ensemble with particle interactions in the hidden sector (see
e.g.~\cite{Chacko:2016kgg,Raveri:2017jto,Hryczuk:2020jhi,Choi:2020pyy})
and  at least $\Gamma_{N-1} \agt H_0$ to combine all of these models. The decays of $X_n$ would also
leave observable imprints on the matter power spectrum~\cite{Dienes:2020bmn}. This
spectrum could play an archaeological role in reconstructing the
properties of the underlying dark sector. A comprehensive study of
the parameter space of such a complex hidden sector is beyond the
scope of this paper and will be presented elsewhere.

\section{Conclusions}
\label{sec:4}

We have studied a set of models endowed with a fraction of CDM
decaying in the early universe to see whether they are able to address
the $H_0$ tension. We have shown that short-lived CDM cannot fully
resolve the $H_0$ tension. The largest contribution to
$\Delta N_{\rm eff}$ in agreement with Planck data corresponds to
$\Delta H_0 \sim 1.3$ and comes from dark matter particles which decay
exclusively into the hidden sector (e.g.  $X \to \psi \bar \psi$),
while leaving completely unmodified the production of light
elements. CDM decaying into photons plus dark radiation is severely
constrained by BBN. It would be interesting to study more complex
models of dynamical dark matter to see whether they can resolve the
$H_0$ tension by combining short-lived and long-lived CDM.

In closing, we comment on relevant aspects of the $^7$Li
discrepancy. Primordial $^7$Li abundance is inferred from observations
of absorption lines in the photospheres of primitive, low-metallicity
stars in the Galactic halo. These objects are warm
($5700 \leq T/{\rm K} \leq 6250$) metal-poor (with small Fe/H
abundances relative to the Sun) dwarf stars. $^7$Li is destroyed in
red giants with core temperatures $T \agt 10^6$~K via the reaction
$^7$Li($p,\alpha$)$^4$He and this is why white dwarfs at moderate
temperatures have been used to deduce the $^7$Li abundance. For each
star, the lithium line strength is used to deduce the Li/H
abundance. Observations from the early 80's appear to indicate that in
low-metallicity stars the $^7$Li abundance is roughly
constant~\cite{Spite:1982dd,Spite:1982nature}. This constant $^7$Li
abundance, usually referred to as the ``Spite plateau'', has been
interpreted as corresponding to the BBN $^7$Li yield. This
interpretation, of course, assumes that lithium has not been depleted
at the surface of these stars so that the presently observed abundance
is supposed to be equal to the primordial one. However, more recent
observations of low-metallicity stars seem to contradict the
conclusions drawn from the Spite plateau. Indeed, the existence of a
metallicity trend in the abundance of $^7$Li in very metal poor stars,
as well as a large dispersion of data have been now observed by many
groups (see e.g.~\cite{Sbordone:2010zi,Melendez:2010kw}) suggesting
that the observed values $^7$Li may not be representative of the
cosmological production
mechanism~\cite{Spite:2012us,Iocco:2012vg}. All in all, it may well be
that the solution of the $^7$Li discrepancy lies inside the stars and
not in cosmology.

\acknowledgments{
I would like to thank Sunny Vagnozzi for discussion. This work has
been supported by the U.S. National Science Foundation (NSF
Grant PHY-1620661) and the National Aeronautics and Space
Administration (NASA Grant 80NSSC18K0464).  Any opinions, findings,
and conclusions or recommendations expressed in this material are
those of the author and do not necessarily reflect the views of the
NSF or NASA.}

\end{document}